\documentclass[fleqn,10pt]{wlscirep}
\usepackage[utf8]{inputenc}
\usepackage[T1]{fontenc}

\usepackage{braket}
\usepackage{graphicx}
\usepackage{xcolor}

\newcommand{\highlight}[1]{{\color{black}#1}}

\def\be{\begin{equation}}
	\def\ee{\end{equation}}
\def\bea{\begin{eqnarray}}
	\def\eea{\end{eqnarray}}

\def\bi{\begin{itemize}}
	\def\ei{\end{itemize}}
\def\ben{\begin{enumerate}}
	\def\een{\end{enumerate}}

\title{Efficient prediction of attosecond two-colour pulses from an X-ray free-electron laser with machine learning}

\author[1,*]{Karim K. Alaa El-Din}
\author[1]{Oliver G. Alexander}
\author[1]{Leszek J. Frasinski}
\author[1,3]{Florian Mintert}
\author[4]{Zhaoheng Guo}
\author[4]{Joseph Duris}
\author[4]{Zhen Zhang}
\author[4]{David B. Cesar}
\author[4]{Paris Franz}	
\author[4]{Taran Driver}	
\author[4]{Peter Walter}
\author[4]{James P. Cryan}
\author[4]{Agostino Marinelli}
\author[1]{Jon P. Marangos}
\author[1,2**]{Rick Mukherjee}

\affil[1]{Blackett Laboratory, Imperial College London, SW7 2AZ, London, UK}
\affil[2]{Center for Optical Quantum Technologies, Department of Physics, University of Hamburg, Luruper Chaussee 149, 22761 Hamburg, Germany}
\affil[3]{Helmholtz-Zentrum Dresden-Rossendorf, Bautzner Landstra{\ss}e 400, 01328 Dresden, Germany}
\affil[4]{SLAC National Accelerator Laboratory, Menlo Park, California 94025, USA}

\affil[*]{karim.alaael-din@physics.ox.ac.uk}
\affil[**]{rick.mukherjee@physnet.uni-hamburg.de}

\begin{abstract}
X-ray free-electron lasers are sources of coherent, high-intensity X-rays with numerous applications in ultra-fast measurements and dynamic structural imaging. Due to the stochastic nature of the self-amplified spontaneous emission process and the difficulty in controlling injection of electrons, output pulses exhibit significant noise and limited temporal coherence. Standard measurement techniques used for characterizing two-coloured X-ray pulses are challenging, as they are either invasive or diagnostically expensive. In this work, we employ machine learning methods such as neural networks and decision trees to predict the central photon energies of pairs of attosecond fundamental and second harmonic pulses using parameters that are easily recorded at the high-repetition rate of a single shot. Using real experimental data, we apply a detailed feature analysis on the input parameters while optimizing the training time of the machine learning methods. Our predictive models are able to make predictions of central photon energy for one of the pulses without measuring the other pulse, thereby leveraging the use of the spectrometer without having to extend its detection window. We anticipate applications in X-ray spectroscopy using XFELs, such as in time-resolved X-ray absorption and photoemission spectroscopy, where improved measurement of input spectra will lead to better experimental outcomes.
\end{abstract}
\begin{document}

\maketitle

\section*{Introduction}
In recent years, X-ray free-electron lasers (XFELs) \cite{Emma, Ishikawa, Allaria} have emerged as a versatile tool for research with applications ranging from damage-free dynamic imaging of molecules\cite{Glownia2016} and proteins \cite{Seibert2011, Pande2016, Chapman2011}, new spectroscopic methods for quantum chemistry \cite{Biggs2013, berrah2011double} and resonant X-ray spectroscopy of nanostructures in condensed matter \cite{Wernet2015, Kroll2018}. The versatility of XFELs is based on their tunability, brightness and very short pulse durations, which make the tracking of ultra-fast dynamics of electrons in matter feasible. 

\begin{figure}[t!] 	
	\centering              
	\includegraphics[width=0.9\linewidth]{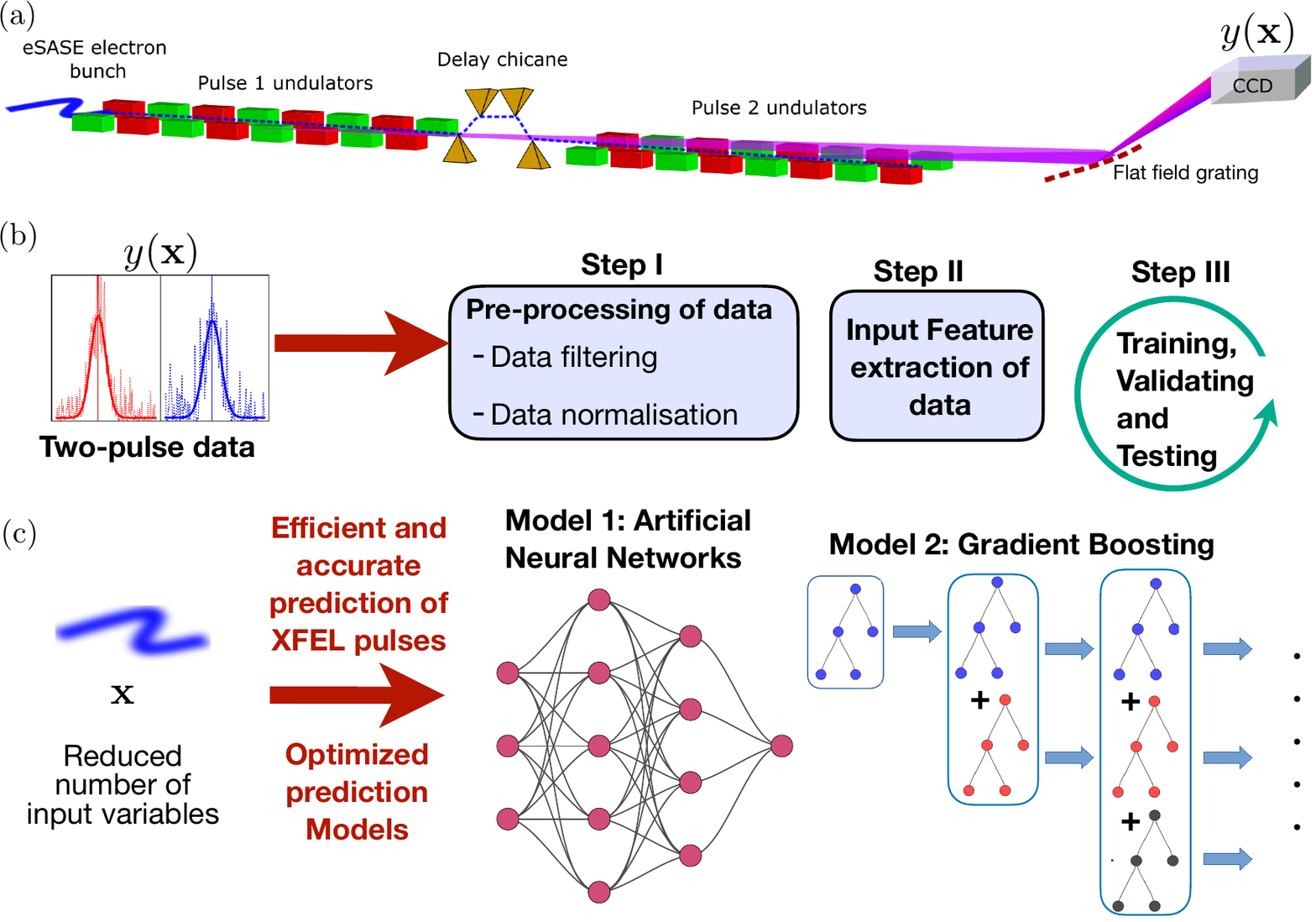}
	\caption{(a) Diagram of the XFEL configuration for two-colour X-ray pulse generation: An electron bunch is modulated in energy-time phase space to yield a high peak current that propagates between two undulator sections separated by a chicane that introduces delay between two pulses. In each undulator section, self-amplified spontaneous emission (SASE) generates a bright, coherent X-ray pulse. A CCD camera is used to measure the spectrum of the two pulses. (b) Diagnostics are used to measure the energies of the two-colour XFEL pulse $y(\mathbf{x})$ which depend on the  input feature vector $\mathbf{x}$. Both $y(\mathbf{x})$ and $\mathbf{x}$ are used to build the prediction model, which consists of three main steps: pre-processing of data, feature extraction, and training/validating/testing of the prediction model. Two different prediction models were used in this work: neural networks and decision tree based on gradient boosting classifier. (c) An optimized neural network or gradient boosting classifier is applied directly to real-time experiments for efficient prediction of central photon energies for  two-colour XFEL pulses.} 
	\label{Fig1}   
\end{figure}

XFEL sources generate X-ray pulses by accelerating electron bunches to relativistic speeds in a linear accelerator of radiofrequency (RF) cavities and allowing them to interact with magnetic fields generated by an undulator\cite{Emma, Ishikawa, Allaria}, see Fig.~\ref{Fig1}. An XFEL can emit coherent or partially coherent radiation because of a favourable self-organization of the electrons in a relativistic beam as it passes through an appropriately tuned undulator. Different configurations are chosen that lead to the modulation of the phase space for the electron bunch and lasing. This  can be used to generate pulses with different properties. Using an additional pre-modulation of the electron beam energy in a short wiggler section, followed by phase space manipulation to transfer the energy into a very short duration high electron current, leads to so-called enhanced SASE that results in sub-femtosecond pulses of the kind studied here \cite{Duris}. SASE and enhanced SASE pulse are important techniques in ultrafast science \cite{Young2018}, where dynamics can be resolved using pump-probe configurations with synchronization to infra-red or optical laser fields \cite{Erk2014, Pande2016} or by using two-pulse XFEL modes \cite{Liekhus-Schmaltz2015,Barillot2021, Picon2016}. Despite the versatility of XFELs in creating two-colour pulses in the femtosecond regime \cite{Lutman}, single-shot variation of the pulse energy is significant; for example, photon energy fluctuation of more than 1\% of the mean, pulse energy up to 100\% of the mean and bandwidth more than 20\% of the mean are common in existing machines. Multiple factors contribute to the instability of output X-ray properties. The working principle of XFEL machines relies on SASE, which is inherently a stochastic process, with amplification seeded broadband emission from noise in the distribution of electrons in the bunch \cite{Bonifacio}. In the case of traditional SASE operation, there are several temporal spikes within the width of the pulse that are not coherent with each other and are amplified, producing only partial longitudinal coherence across the XFEL pulse. This is compounded by fluctuations in the RF amplitudes or RF phases, which can translate to variation of the spatial and energy distribution of the electrons within a bunch.

Techniques like XFEL seeding and optical active stabilization may improve stability, but the issue of temporal fluctuations is still relevant at the few-femtosecond level. Alternatively, one can also circumvent issues of  unstable  pulse properties by performing a full X-ray characterization for each XFEL shot. However, single-shot characterization of XFEL pulses requires higher-dimensional inputs, such as the X-ray spectrum,  which are obtained in a data expensive manner e.g. using an X-ray spectrometer with a CCD image readout. In addition to the slow and invasive diagnostics, the processing of large volumes of image data, given inevitable limits to computational power and data transfer rates, restricts the rate of characterization \cite{Ding,Harmand,Kimberg}. Diagnostics in current machines operate at kHz repetition rates, and technological advances in high speed diagnostics must be accompanied by increased efficiency to reduce complexity and cost. An interesting solution to the issue of slow characterization of XFEL pulses was suggested in \cite{Sanchez-Gonzalez}, where machine learning techniques were used to make accurate predictions of  XFEL properties using data collected solely from fast diagnostics. The key concept relies on exploiting the correlation of various XFEL properties such as photon energy and spectral shape of the X-ray pulses with data that can be acquired at a higher repetition rate, such as electron beam properties. Since the detailed modelling of every experimental aspect that determines this correlation is currently out of reach, machine learning methods can prove to be extremely useful in this context, as further illustrated in \cite{ren2020temporal}. Whilst the quantum fluctuations associated with SASE will not be amenable in principle to machine learning, the complex interplay of the other fluctuating parameters gives some hope that machine learning strategies can be applied to predict the X-ray parameters with improved fidelity. 

In this work, we use  techniques of supervised learning to make efficient predictions of central photon energies for attosecond fundamental and harmonic pulses with high fidelity that can be applied to any XFEL facility. Enhanced SASE is realised by manipulation of the electron bunch spikes from the photoinjector with the undulator split into two sections for radiation of $\omega$ and $2\omega$ frequencies \cite{Guo2023}. 
\highlight{We use two different approaches of supervised learning, namely artificial neural networks (ANNs) and gradient boosted decision trees (GB) for our predictions. While the former consists of multiple layers with inter-connected nodes (artificial neurons), the latter constitutes of an ensemble of decision trees with better performance and lower overfitting than simple decision trees.  By applying feature selection analysis, we reduce the dimensionality of the entire input space to the most relevant features.} This leads to a simpler neural network architecture and optimal decision trees that make accurate predictions for real experimental data while enhancing the training efficiency when compared to \cite{Sanchez-Gonzalez}. Moreover, despite XFEL beamlines being typically designed with the flexibility to allow for different experimental configurations (targets, diagnostics, etc.), at current facility beamlines it is not usually possible to measure the X-ray spectrum before and after a sample. Many experiments are also unable to measure multiple pulses simultaneously, due to the limited spectral range of available spectrometers. One of the key results of our work is the intriguing possibility of using machine learning methods to predict the photon energy for the second harmonic pulse without relying on the measurements of the fundamental pulse. Thus our methods offer a more pragmatic approach to maximising useful information from available resources whilst adding little experimental overhead.  

\section*{Building the prediction model}
A prediction model mathematically connects the output variables to the input parameters. This mathematical function is often non-trivial, especially for noisy experiments, which exhibit large variance of the affected parameters and variables. This leads to difficulties in discriminating between noise and signal, while further establishing an upper bound on the quality of predictions we can achieve. Naturally, the quality of the model is benchmarked by its ability to make successful predictions for future measurements.  

Figure~\ref{Fig1} illustrates machine learning of the prediction model. The objective is to predict the pulse characterization y from the diagnostics x. There are three main stages to building the theoretical prediction model. The first step is to perform pre-processing on the raw experimental data, which mainly involves filtering and normalizing the data. Here, filtering implies removing outlier events, such as events that correspond to low variance or based on not properly recorded measurements. The next step is to randomly split the pre-processed data into three different data sets: $70\%$ of the data set used for training to fit different models, $15\%$ of the data set used for testing while another $15\%$ used for validation. The models chosen for this work are artificial neural networks \highlight{(ANNs)}\cite{BingCheng} and gradient boosting \highlight{(GB)} \cite{Chen}. We train, validate and benchmark the performance of the prediction models on the test set. Later in this work, the performance of the machine learning methods are compared with a simpler model, namely a linear regression model \cite{Schneider}. The final step is to optimize the prediction model in terms of its training cycle period. For this, it is important to identify the most relevant input features that contribute to the prediction of the pulse properties, especially since an unnecessarily large number of input features can slow down the fitting of estimators as well as decrease the quality of model predictions by over-fitting. The reduced input space leads to a simpler and more robust prediction model.

\section*{Results}

\subsection*{Reducing the dimensionality of input space} 
The goal is to identify the most relevant set of input features, which in this case are the XFEL electron beam properties, by assessing their importance in the prediction of the output. Typically, a few hundred parameters are recorded for each event, including measurements of the electron beam properties, basic photon diagnostics (such as gas detectors for the pulse energy) and large numbers of other environmental variables. Many of the environmental features are collected at a reduced rate of 1 Hz and therefore are only measuring slower fluctuations. This is done to reduce data flow rates, as these variables are generally uninformative at high repetition rates but could, in principle, be measured on every shot. A lot of these parameters such as environmental variables are empirically known to be disconnected from the XFEL operation and thus have no predictive value. These are systematically removed to reduce the total number of input features for an event from hundreds to $N\simeq80$. Focusing on the remaining features, especially with those that have large fluctuations, it is a priori unclear whether they are expected to have predictive value. For such instances, it is useful to perform a thorough statistical analysis on the remaining features and rank them  in order of their relevance using the permutation feature importance function \cite{Breiman}.
\begin{figure}[ht] 	
	\centering        
    \includegraphics[width=0.8\linewidth]{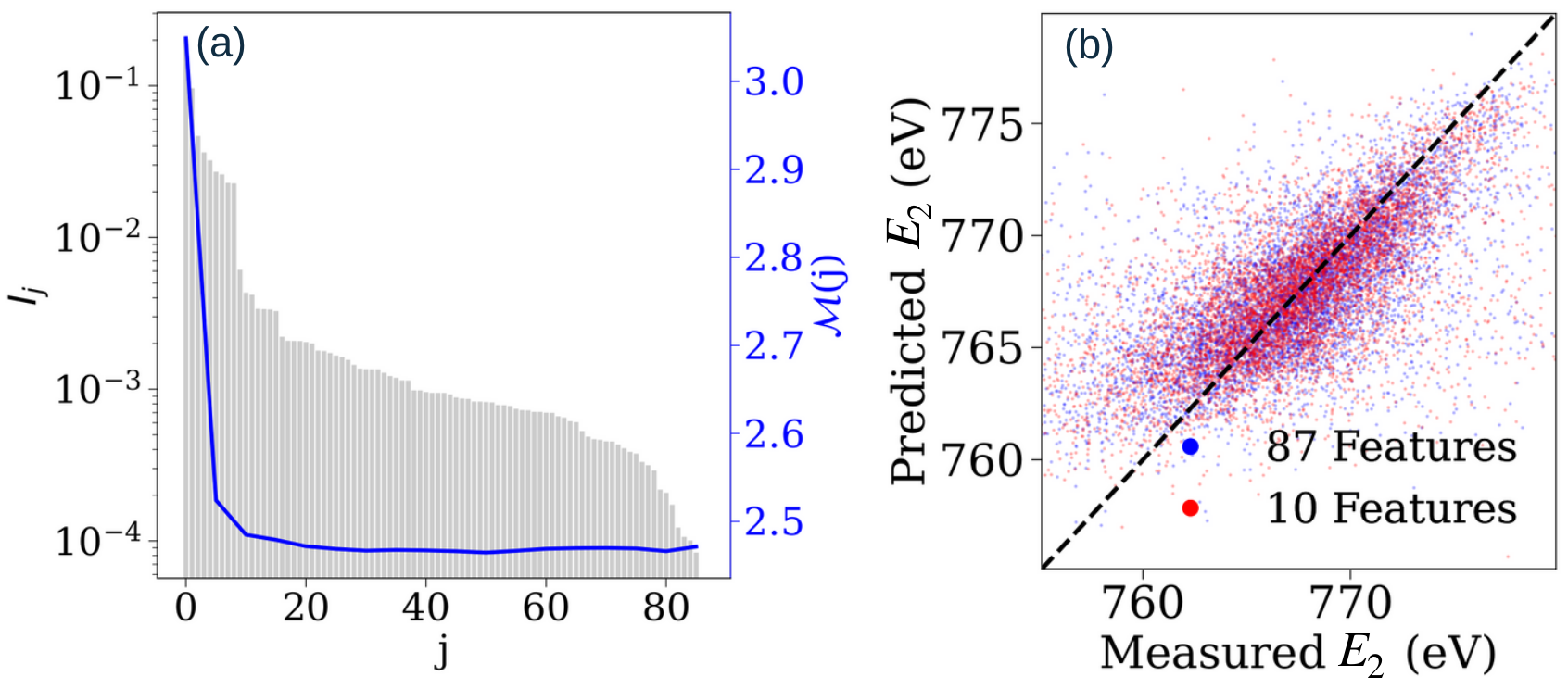} 
    \caption{\highlight{The grey bars in }panel (a) depict the permutation importance function $I_j$ for a particular input parameter $j$ while predicting the central photon energy of the second pulse ($E_2$) using neural networks. Mean absolute error $\mathcal{M}$ (solid blue line, in $eV$) is plotted for a varying number of input features \highlight{selected by feature importance}. Panel (b) is a scatter plot that compares the measured values of $E_2$ with the predicted values of neural networks. Predictions of the reduced input space (red dots) agree with the full input space (blue dots) with a mean absolute error of $\mathcal{M}=2.48$ eV.} 
	\label{Fig2_features}   
\end{figure}
Before describing the importance function, we define the input matrix denoted by $\tilde{\mathbf{x}}$ whose dimensions are $S$ (total number of events) $\times$ $N$ (input features for each event). Throughout this work, the tilde will be used to indicate that the data have been normalized to zero mean and unit standard deviation. Thus, for $i^{\text{th}}$ event, the row vector has $N$ input features denoted by the vector $\mathbf{\tilde{x}}_i = (\tilde{x}^{1}_i, \tilde{x}^{2}_i, \hdots, \tilde{x}^{N}_i)$ while for the $j^{\text{th}}$ input feature, the column vector has $S$ events denoted by $\tilde{\mathbf{x}}^{j} = (\tilde{x}^{j}_1, \tilde{x}^{j}_2, \hdots,\tilde{x}^{j}_S)^\mathsf{T}$. The mean absolute error calculated over $S$ events is given as
\be
\mathcal{M}(\tilde{\mathbf{x}}, N) = \frac{1}{S}\sum\limits^S_{i=1}|\tilde{Y}_i - f(\tilde{\mathbf{x}}_i, N)| ,
\ee 
where $\tilde{Y}_i$ denotes the output for the $i^{\text{th}}$ event and $f(\tilde{\mathbf{x}}_i,N)$ is the estimator for the output observable generated using the input vector $\mathbf{\tilde{x}}_i$. The relevance of a particular $j^{\text{th}}$ input feature is given using the normalized permutation feature importance function \cite{Breiman} which is denoted here by $I_j$. It measures the increase in the mean absolute error when the $j^{\text{th}}$ input feature is randomly replaced by an incorrect one and is defined as follows,
\be\label{Imp}
I_j = \frac{1}{\mathcal{M}(\tilde{\mathbf{x}}, N) }\left(\frac{1}{R}\sum\limits^R_{r=1} \mathcal{M}(\mathbf{p}^{r}(j), N) - \mathcal{M}(\tilde{\mathbf{x}}, N) \right),
\ee
where $\mathbf{p}^r(j) = (\mathbf{p}_1^{r}(j), \mathbf{p}_2^{r}(j), \hdots, \mathbf{p}_S^{r}(j))^\mathsf{T}$ is a matrix of the $r^{\text{th}}$ permutation to the $j^{\text{th}}$ input feature. Its individual row vectors are denoted as $\mathbf{p}_i^{r}(j) = (p^{1,r}_{i}(j), p^{2,r}_{i}(j) \hdots, p^{N,r}_{i}(j))$. These vectors have elements where only the $j^{\text{th}}$ input feature is replaced using a permutation operator $\Pi^r$ which gives the element
\be
p_{i}^{k, r}(j) =
\begin{cases}
	\tilde{x}_i^{k} &\quad\text{if}\quad k\neq j,\\
	[\Pi^r(\tilde{\mathbf{x}}^{j})]_i &\quad\text{if}\quad k= j.\\
\end{cases}
\ee
Here $\Pi^r(\tilde{\mathbf{x}}^{j})$ gives the $r^{\text{th}}$ permutation from a series of random permutations applied to column vector $\tilde{\mathbf{x}}^{j}$. The $i^{\text{th}}$ value of the resultant vector obtained after the permutation is given by the element $[\Pi^r(\tilde{\mathbf{x}}^{j})]_i$. All other column vectors $\tilde{\mathbf{x}}^{k \neq j}$ remain unaltered.

Figure~\ref{Fig2_features}(a) is a plot which ranks the input features using the permutation feature importance $I_j$ while predicting the central photon energy of the second pulse $E_2$ \highlight{with an ANN} using only non-pulse measurement data. The relevance of a particular input feature is ranked with descending values of $j$ and the plot of the mean absolute error ($\mathcal{M}(j)=\mathcal{M}(\tilde{\mathbf{x}}, j)$) reaches its lowest value for the top ten relevant features, most of which are related to the electron beam properties. A listing with descriptions of the ten most important features is given in table \ref{tab1:imp}. Adding further features leads to over-fitting, as is seen with the rise in $\mathcal{M}(j)$ for higher $j$ values. 

Figure~\ref{Fig2_features}(b) shows a scatter plot which compares the measured values of the central photon energy of the pulse $E_2$ with the predicted values estimated by the ANN. \highlight{The predictions obtained with GB match these both quantitatively and qualitatively, as illustrated for a range of data in the supplemental.} For a perfect predictor, the points would all lie exactly along the diagonal, with deviations from this distribution indicating reduced accuracy of prediction. The blue and red scatter points correspond to full input space ($M=N=87$) and reduced input space ($M=10$) respectively. \highlight{The main deviations in this prediction are shared between both full and reduced input spaces, and are visible as the weak nearly uncorrelated background, and the deviation of the predictions far from the mean energy. The former is likely due to the highly stochastic nature of SASE, while the latter is indicative of low estimator confidence leading to more conservatice estimates closer to the mean. These error signatures are nearly identical for both full and reduced spaces, and} the overall quality of predictions was identical, with a mean absolute error of 2.48 eV. Thus, we can perform training of simpler estimators with smaller architectures by using the reduced input space, without compromising the quality of predictions. By including only the  most relevant features, we introduce a feature-restricted mean absolute error $\mathcal{M}=\mathcal{M}(\tilde{\mathbf{x}}, 10)$, which will be used to estimate the performance of predictor models for the rest of this work. To further allow for comparability between different prediction targets, we will proceed by normalizing the mean absolute error $\mathcal{M}$ with respect to the standard deviation $\sigma$ of the target data. Using this notation, the results seen in Fig.~\ref{Fig2_features}(b) are equivalent to $\mathcal{M}=0.54\sigma$. Whilst the accuracy of the predictions is modest, this was achieved without addition probes of electron and X-ray properties to those already in use at LCLS. The methods employed in this work can \highlight{therefore} be used generally for the prediction of beam properties. For example, the results shown in Figs.~(S1) and (S2) of the Supplemental Material were obtained using a completely different experimental setup  \cite{Sanchez-Gonzalez} and provide $\mathcal{M}\sim0.2\sigma$ and $\mathcal{M}\sim0.3\sigma$ for the prediction of the time delay and central energies respectively. The data for Figs.~(S1) and (S2) indicate that the input-output correlation in the data for the time delay parameter between the pulses is much higher than that of the central photon energies of the pulses. \highlight{The difference in performance between predictions for these two experiments indicates that the limiting factors are the specifics of experimental setup and inherent noise, rather than the machine learning method itself. These limitations likely manifest themselves in low correlation between input features and labels, and errors in the ground-truth of measurements, respectively.}

\subsection*{Independent prediction of a single pulse} 
\begin{figure}[tb!] 	
	\centering              
	\includegraphics[width=0.8\linewidth]{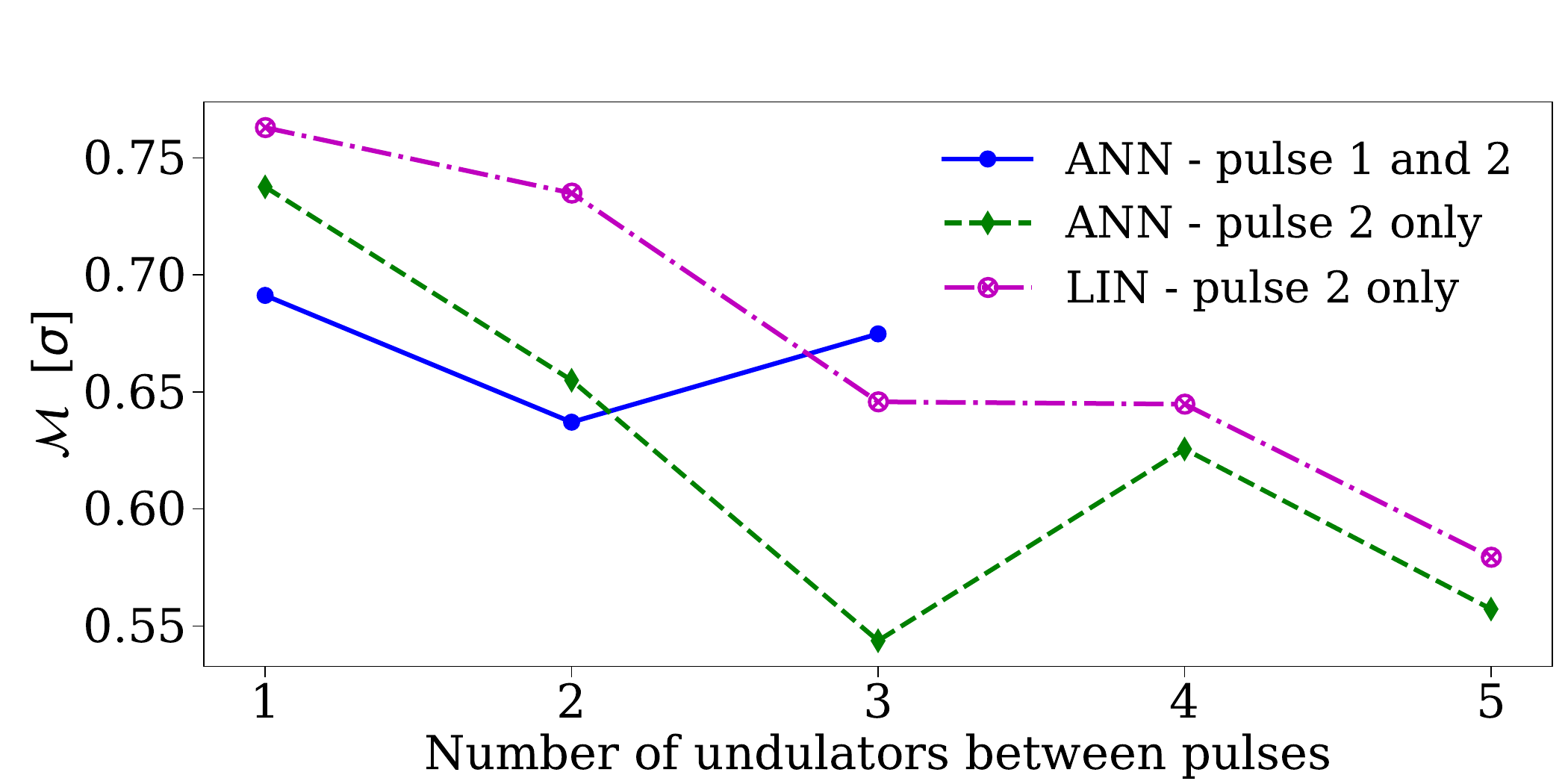} 
	\caption{The plot shows the precision with which the central photon energy of the second pulse ($E_2$) can be predicted as a function of the number of undulators between the pulses for two different detection scenarios: one where both pulses are measured  simultaneously, while for the other only pulse 2 is measured. Using the former data, ANN is used to predict $E_2$ (blue solid line with circle), while using the latter data, both LIN (magenta dashed line with circle) and ANN (green dashed line with triangle) were used to predict $E_2$.} 
	\label{Fig4:pulse2}   
\end{figure}

Figure~\ref{Fig4:pulse2} focuses on predicting the central photon energy of Pulse 2 ($E_2$) using  two different detection schemes for the experiment. One setting corresponds to a configuration of the spectrometer which detects both the pulses simultaneously (depicted with blue lines) and using the energy of Pulse 1 ($E_1$) as an input feature, while the other measures only the second pulse (depicted with dashed lines). The green dashed line is the prediction of the central photon energy for Pulse 2 with ANN, while the magenta dashed line is with LIN model. These predictions are made with experimental data where different numbers of undulators were used between the pulses. Although both LIN and ANN models make accurate predictions of $E_2$ without the spectral information of Pulse 1, we find that the accuracy of their predictions depends on the number of undulators between the pulses. Predictions of $E_2$ improve with increasing number of undulators that are used for generating the second pulse. 
One plausible explanation for this is that, as each additional undulator provides amplification to Pulse 2, the accuracy of central-photon energy \highlight{estimation} (the ground truth of our prediction models) improves. \highlight{Alternatively, this can be understood by considering the interaction times each pulse shares with the electron beam. The first pulse only shares a short interaction with the beam, so it may not correlate with properties of the entire beam, butresults rather only a part of it. The second pulse is seeded by the first, which leads to high correlation between the two and similarly bad predictability of the second pulse for low undulator counts. However, as the number of undulators for the second pulse goes up, its interaction time with the beam increases and this may explain the improved predictions using overall beam properties.}


It \highlight{is further worth noting} that although Pulse 2 is a harmonic of Pulse 1 and is generated from the same electron bunch, the spectrometer was optimized for Pulse 2 and thus the accuracy in determining $E_2$ for both the training and test data is improved when compared to the setup where both pulses were measured. Often in experiments, measurement of the energy spectrum of both pulses simultaneously is not possible due to the limited spectral range of the spectrometer. Furthermore, it may only be possible to measure photon spectra after transmission through target samples, which in many settings alter the spectrum, e.g. due to absorption. This result allows for prediction of the photon energy without input from the spectrometer, except for training, adds directly to the capabilities of current XFEL experiments, allowing for important information about the incoming pulses to be extracted within typical experimental constraints.

\section*{Discussion}
Conventional X-ray spectrometers involve high volumes of data and are still too slow for future XFEL experiments (which will run at MHz repetition rates) and proposed high data rate models using photo-electron spectrometers \cite{Li2022a, Heider2019} would add significantly to experimental cost and complexity. Another issue is the limited spectral range of the available spectrometers. In both cases, machine learning methods can be advantageous, as demonstrated in this work. Although there have been prior works relying on the concept of using data from the photon spectrometer to train the neural networks, our work suggests that gradient boosting methods are more efficient (orders of magnitude) than neural networks in making spectroscopic predictions  while giving comparable predictions. It is well-established that there is strong dependence of the properties of two-colour pulses on the electron beam parameters. \highlight{Although} most of the environmental variables are usually not relevant, it could be that certain environmental parameters specific to that facility beamline play a crucial role in making more accurate predictions. One of the challenges in pre-processing of data used for predictor models is to filter the relevant features from the redundant ones. In this work, the dimension of the input parameter space was drastically reduced without having to compromise with the prediction results using the feature selection analysis. However, the data collected in the experiment was not tailored to machine learning, and the electron beam and photon properties recorded were incidentally of use for predictions and, in future experiments, collection of more relevant electron beam properties may allow for improved prediction accuracy.

\section*{Methods}
\subsection*{A: Experiment details for attosecond two-colour pulses}
In our experiment, data with two pulses at different energies were obtained from a configuration similar to \cite{Duris}, utilizing an enhanced SASE mode. The phases between SASE emitting microbunches are not predetermined and, as a result, the temporal properties are difficult to predict from purely spectral measurements. The photon energy of the emission is determined by the period of the undulators, the energy of the electron bunch and the position of the SASE emission within the bunch. The spatial and energetic distribution of electrons within the bunch varies on a shot-to-shot basis due to fluctuations in the electron accelerator. In the two-colour mode, a second set of undulators was used to produce a second pulse (see Fig. \ref{Fig1}), either at the second or third harmonic of the first, with the emission from the first pulse seeding the second. 

Separation of the X-rays from the electrons due to a difference in their group velocities, i.e. slippage, was used to create a time delay between the pulses, for use in a separate pump-probe experiment. With more undulators in the second section, the slippage is larger at the centre of mass of the second pulse generation, so the delay is greater. Both pulses are estimated to have temporal length below 500 attoseconds \cite{Duris}. Pulses were generated at 120 Hz with photon energies of approximately 250 eV and using either the second or third harmonics at 500 and 750 eV respectively,  2--10 eV FWHM bandwidth, and up to 50 $\mu$J energy in each pulse.

\subsection*{B: Pre-processing of data}
\subsubsection*{Data Filtering}
A typical experimental data set will contain many events which are labelled by $i\in1,2\hdots,S$, where $S=35000\text{--}40000$. After filtering, the total number of events in each data set reduces to $S=16000\text{--}32000$ (varying between the different data sets) that can actually be used for building the predictive model. For each event, we typically have around 300 recorded input features that are collected during the experiment. These include environmental variables such as current and voltage measures for different XFEL machines, total photon energies of the pulses as measured by gas monitor detectors as well as electron beam properties at the dump which include electron beam charge and energy.
We remove from this set of features any that take less than $10$ distinct values across the full dataset. Furthermore, we eliminate any features that are perfectly correlated (correlation coefficient above 0.995). The combination of these two methods brings our overall feature count down to around 80 (depending on the individual data set). Based on the statistical dispersion of the data, we also remove \textit{outlier events} which can negatively impact the prediction results. Thus, any events with features with a median absolute deviation greater than four are removed. Finally, we impose a lower limit of $5~\mu$J on the total central photon energy of the pulse as measured by the gas monitor detectors.

\subsubsection*{Normalisation of data}
Let the vector of input features for event $i$ be denoted by $\mathbf{x}_i = (x^{1}_i, x^{2}_i, \hdots, x^{N}_i)$  where $N$ is the total number of recorded features and the output for this event be denoted by $Y_i$. Then the normalised input and output data are given as
\bea
\tilde{x}^{j}_i &=& (x^{j}_i - \mu_{\mathbf{x}^{j}}) / \sigma_{\mathbf{x}^{j}} \nonumber \\
\tilde{Y}_i &=& (Y_i - \mu_\mathbf{Y}) / \sigma_\mathbf{Y}
\eea
Here $\mathbf{x}^{j}=  (x^{j}_1, x^{j}_2 \hdots, x^{j}_S)^\mathsf{T}$ is a vector consisting of $j$th input variable from every event and $\mathbf{Y} = (Y_1, Y_2\hdots, Y_S)^\mathsf{T}$ is the output vector. Additionally, $\mu$ and $\sigma$ respectively correspond to the mean and standard deviation of the subscripted data column across all events.

\subsection*{C: Key Code for top ten input parameters for Fig.~2(a)}
\begin{table}[!ht]
	\centering
	\begin{tabular}{|p{0.75cm}|p{4cm}|p{10cm}|p{1cm}|}
		\hline
		\textbf{Rank} & \textbf{Label}        & \textbf{Description}        & \textbf{Data rate}                                                                                                                             \\ \hline
		1    & ebeam\_ebeamLTU450       & \highlight{A position based energy measurement of the electron beam orbit in a dispersive region of the linac-to-undulator beamline}       & 120 Hz                            \\ \hline
	 	2	 & ebeam\_ebeamLTU250       & \highlight{A position based energy measurement of the electron beam orbit in a dispersive region of the linac-to-undulator beamline}       & 120 Hz                            \\ \hline
		3    & epics\_UND\_34\_gap\_act & The measured vertical gap between the undulator magnet arrays in one undulator, which is tuned to adjust the K-factor of the undulator       & 1 Hz \\ \hline		
		4    & epics\_UND\_36\_gap\_act & The measured vertical gap between the undulator magnet arrays in another undulator       & 1 Hz                                                   \\ \hline		
		5    & xgmd\_energy             & The pulse energy measured after all attenuation using the total ionisation of an $N_2$ filled cell (X-ray gas monitor detector)       & 120 Hz       \\ \hline
        6 & epics\_UND\_34\_gap\_des    & The desired vertical gap between the undulator magnet arrays in the first undulator       & 1 Hz                                                   \\ \hline
		7 	 & ebeam\_ebeamPkCurrBC2    & The peak electron bunch current measured in the second bunch compressor       & 120 Hz                                                    \\ \hline	
		8    & epics\_GMD\_ElectronMesh        & The voltage applied to an electron mesh in the gas monitor detector (GMD), to extract the electrons towards the electrode       & 1 Hz  		 \\ \hline
		9    & gmd\_energy & The pulse energy measured after all attenuation using the total ionisation of an Kr filled cell (X-ray gas monitor detector)       & 120 Hz   \\ \hline 
		10  & epics\_UND\_28\_gap\_act & The measured vertical gap between the undulator magnet arrays in a third undulator       & 1 Hz                                                   \\ \hline                                   
	\end{tabular}
	\caption{Input feature ranking by permutation feature importance for pulse energy data}
    \label{tab1:imp}
\end{table}


\subsection*{D: ML methods}
\subsubsection*{Linear modeling}
A linear regression model (LIN) fits a general linear function
\be
\Bar{Y}^{(LIN)}_i = \tilde{\mathbf{x}}_i\cdot\mathbf{c}+c_0
\ee
across $S$ events. The parameters $\mathbf{c}, c_0$ are varied to minimize the residuals-squared, given by
\be
RS = \frac{\sum\limits^S_{i=1}(\tilde{Y}_i-\Bar{Y}_i)^2}{\sum\limits^S_{i=1}(\tilde{Y}_i-\tilde{\mu}_Y)^2}.
\ee
Here, $\tilde{\mu}_Y$ is the mean of the normalized labels $\tilde{Y}$ such that $\tilde{\mu}_Y \equiv 0$. We then use the mean absolute error $\mathcal{M}$ to calculate the model performance. While linear regression methods can be very useful and simple to implement, they naturally fail with data that are highly non-linear. Since the generation of XFEL pulses are highly non-linear processes, it is helpful to use this method to get a sense of  the level of non-linearity in the data set.

\subsubsection*{Gradient boosting decision trees}
Decision tree learning is a supervised machine learning approach often used for predicting classification or regression type of problems. A decision tree is built by splitting the root node (which is at the apex) into subsets, and this process of splitting continues for each subset recursively until further splitting does not improve the predictions. The rules for splitting a node are determined by the classification features. Gradient boosted decision trees is an ensemble learning method where rather than using a single decision tree to make predictions, we combine multiple decision trees to enhance the model's accuracy. The basic premise of boosting is to combine weak ``learners" into a single strong learner iteratively. The success of the boosting scheme is evaluated by defining a suitable loss function that is minimized using a gradient descent scheme.

\begin{figure}[htb!] 	
	\centering              
	\includegraphics[width=0.8\linewidth]{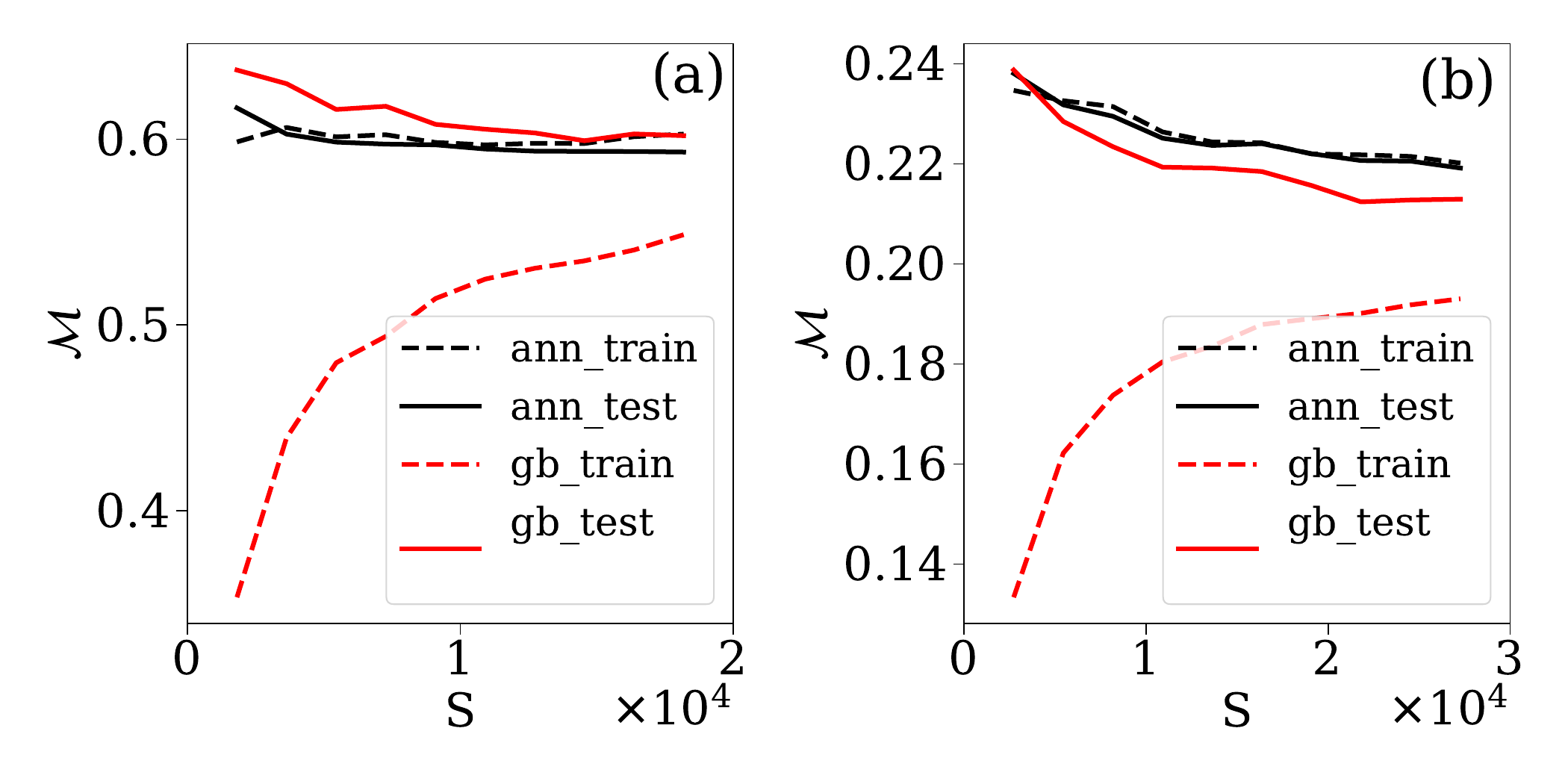} 
	\caption{Convergence of the mean absolute error as function of the number of events S used for training the decision trees/neural networks for (a) Prediction of central photon energy of the pulse using \cite{Duris} and (b) time delay prediction using \cite{Sanchez-Gonzalez}.} 
	\label{Fig5_samples}   
\end{figure}

In our case, the full set of events $S$ forms the root node, which is subsequently split into subsets $\mathcal{S}_i$ and are distinguished based on the values of different categorical or numerical features. We partition out the input space into $D$ regions $d\in1,\hdots,D$ where we split the data using
\be
z_d(\tilde{\mathbf{x}}_i) = \begin{cases}
	1  &\quad\text{if}\quad \tilde{\mathbf{x}}_i\in d,\\
	0  &\quad\text{if}\quad \tilde{\mathbf{x}}_i\notin d.\\
\end{cases}
\ee
By predicting a constant value $h_d$ across each of these regions, we can define the output of the decision tree as
\be
\Bar{y}_t(\tilde{\mathbf{x}}_i; N) = \sum\limits_{d=1}^Dh_d z_d(\tilde{\mathbf{x}}_i).
\ee
Here, $h_d$ is the average of the target across all points within the region $d$, and is used as the model output for all points where $z_d=1$. The predictions of an individual decision tree are generally heavily biased, and thus ensemble methods are often used. Apart from random forests, which use independent decision tree predictors, gradient boosting (GB) is another commonly used method where trees are added to the estimator successively and fitted to the pseudo-residuals of all the previous tree's predictions. A gradient boosting regressor\cite{Chen} is an ensemble method that gives an estimate $\Bar{Y}^{(GB)}_i$ from the weighted sum of estimates given by $T$ base regressors $\Bar{y}_t(\tilde{\mathbf{x}}_i; N)$, written as

\be
\Bar{Y}^{(GB)}_i = \sum\limits_{t=1}^T\gamma_t \Bar{y}_t(\tilde{\mathbf{x}}_i; N),
\ee
where we used the decision trees to define our base estimator. The gradient boosting regressor is then constructed iteratively under consideration of a differentiable loss function
\be
\mathcal{L}=\frac{1}{S}\sum\limits^S_{i=1}(\tilde{Y}_i - \Bar{Y}_i)^2.
\ee
We begin by considering a constant average estimate $\Bar{Y}^{(GB)}_{i, 0} = \tilde{\mu}_Y=0$, where the subscript 0 indicates that no estimators have been added yet. We then iterate over $t\in1,\hdots,T$ and at each step perform the following:
\begin{enumerate}
	\item For each $i$, find the pseudo-residuals given by
	\be
	q_{i,t} = - \frac{\partial\mathcal{L}}{\partial\Bar{Y}_{i, t-1}}.
	\ee
	\item Fit a decision tree estimator $y_t(\tilde{\mathbf{x}}_i; N)$ to the set of pseudo-residuals.
	\item Find $\gamma_t$ to minimize $\mathcal{L}$ for the new set of estimates
	\be
	\Bar{Y}_{i, t} = \Bar{Y}_{i, t-1} + \gamma_t y_t(\tilde{\mathbf{x}}_i).
	\ee
\end{enumerate}
After adding $T$ base estimators in this manner, we have our fully fitted estimator $\Bar{Y}^{(GB)}_i = \Bar{Y}_{i,T}$. This approach has the advantage of focusing on regions of bad prediction and improving them. While many tree parameters are fit in the algorithm, others are hyperparameters that have to be specified a priory, such as the number of trees, the number of decisions per tree, the use of regularization and the number of data points to consider for each decision. Often the intuitive interpretation of the regressor obtained from decision trees can be lost when using an ensemble of decision trees. We found an estimator with 20 trees without specified depth limit and l2 regularization to yield the best results with only minor overfitting as seen in Figure~\ref{Fig5_samples}. To evaluate the performance of the gradient boosting estimator, we evaluated the mean absolute error across the test set and compare it to the performance of the ANN and the linear model.

\begin{figure}[htb!] 	
	\centering              
	\includegraphics[width=0.8\linewidth]{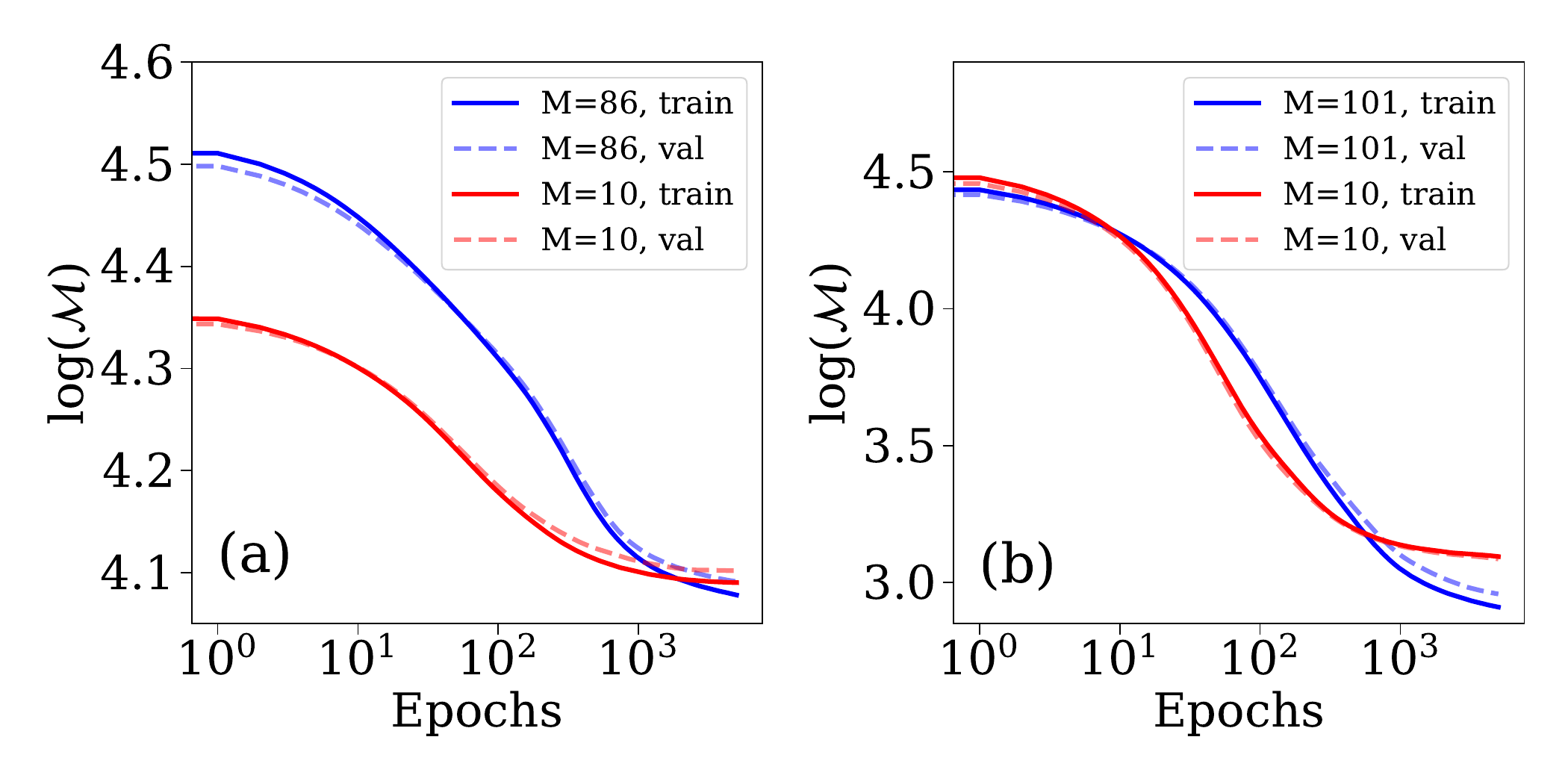} 
	\caption{Convergence of mean absolute error as function of the number of epochs used in the neural networks for (a) predicting the central photon energy of the pulse using \cite{Duris} and (b) time delay prediction using \cite{Sanchez-Gonzalez}.} 
	\label{Fig6_epochs}   
\end{figure}

\subsubsection*{Neural networks}
Artificial Neural Networks (ANNs) are one of the most widely used modern machine learning techniques and have been very successful in making predictions for various physical systems. In this work, we use Feed-Forward Neural Networks as we are performing supervised learning on a set of independent data points. Conceptually, a neural network can be represented by a graph, with values and biases associated with each node (or neuron) and weights associated with each edge. We group the nodes into layers, and allow edges only between nodes of neighbouring layers. The data propagates through this network layer by layer in one direction (Feed-Forward) only. The overall architecture of the neural network is defined by the hyperparameters which include the number of neurons in each layer, number of layers and choice of activation function applied to the outputs of different nodes. Regularization schemes and choice of optimizer constitute further hyperparameters, while bias $b$ and weights $W$ are parameters fit using the backpropagation algorithm. The last layer must have the same size as the number of prediction labels in the data, 1 in our case. For each of the $L+1$ layers labelled by $l\in0,\hdots,L$, we define the node activation by a vector $\mathbf{v}_l$, the node bias by a vector $\mathbf{b}_l$, the edge weights for edges between layers $l$ and $l+1$ by a matrix $\mathbf{W}_l$ and the differentiable activation function for each node in the layer as $a_l$. We then perform forward propagation of the data for event $i$ by setting $\mathbf{v}^{i}_0 = \tilde{\mathbf{x}}_i$. We then propagate the data using
\be
\mathbf{v}^{i}_{l+1} = a_l\left(\mathbf{W}_l\mathbf{v}^{i}_l+\mathbf{b}_l\right)
\ee
and use $\Bar{Y}^{(ANN)}_i=\mathbf{v}^i_L$ as our estimate of $\tilde{Y}_i$. The crucial task is then to train the estimator by finding $\mathbf{W}_l$ and $\mathbf{b}_l$ such that our loss, chosen as $\mathcal{M}$ is minimized. We initialize these parameters randomly, and then perform backpropagation with gradient descent, implemented through the Adagrad algorithm \cite{pedregosa2011scikit}. We used Bayesian optimization to find the optimal neural network architecture, activation functions, regularization and drop out. This technique uses Bayesian inference to guess combinations of hyperparameters that yield the best predictions for the smallest computational cost. We find that the optimal network sufficient to make accurate predictions for both the two-pulse delay and the pump-probe energies consists of two hidden layers of 20 cells each. The network is also l2-regularized and there is no drop-out, leading to no overfitting (Figure~\ref{Fig5_samples}) and training convergence after few thousand of epochs (Figure~\ref{Fig6_epochs}). The choice of the activation function on hidden layers is chosen to be a ReLU (regularized linear unit function). In combination with the reduced feature count, this results in a substantial speed-up of model fitting and requires far fewer data to be collected.

\section*{Data availability}
The raw data for this research was generated at the Linear Coherent Light Source, both raw and processed datasets are available upon reasonable request to the corresponding author.

\section*{Code availability}
The codes used for this work are available upon reasonable request to the corresponding author.

\section*{Author contributions statement}
The project was conceived by KKA and RM. KKA performed the machine learning and data analysis guided by RM. The experimental data was provided by ZG, JD, ZZ, DBC, PF, TD, PW, AM, JPC, and JPM, while OGA did the pre-processing of the data. RM, KKA, OGA, LJF, FM, JPM and JPC contributed to the writing of the manuscript. 

\section*{Acknowledgements}
JPM would like to acknowledge EPSRC funding EP/X026094/1. AM would like to acknowledge support from US Department of Energy (DOE), BES Scientific User Facilities Division Field Work Proposal 100317; JD and AM were supported by the Laboratory Directed Research and Development Program in support of the Panofsky fellowship. The contributions from TD and JPC were supported by the US DOE, Office of Science, Office of Basic Energy Sciences (BES), Chemical Sciences, Geosciences, and Biosciences Division (CSGB).  Use of the Linac Coherent Light Source (LCLS), SLAC National Accelerator Laboratory, is supported by the US DOE, Office of Science, BES, under Contract DE-AC02-76SF00515.

\section*{Competing Interests}
The authors declare no competing interests.

\section*{Supplemental Material}

\setcounter{equation}{0}            
\setcounter{section}{0}    
\setcounter{figure}{0}    
\renewcommand\thesection{\Roman{section}.}    
\renewcommand\thesubsection{\arabic{subsection}}    
\renewcommand{\thetable}{S\arabic{table}}
\renewcommand{\theequation}{S\arabic{equation}}
\renewcommand{\thefigure}{S\arabic{figure}}
\setcounter{secnumdepth}{2}	

\noindent{} Here we present the results of our ML prediction for central photon energies using the experimental setup \cite{Duris} described in the main text and compare it against the predictions of energies and time delay using data obtained from another two-colour experiment \cite{Sanchez-Gonzalez} which has a different modus operandi than \cite{Duris}. Although \cite{Sanchez-Gonzalez} does not utilize an enhanced SASE scheme like in \cite{Duris}, both methods use a variable line spectrometer to measure the X-ray spectrum. To create the two pulses, a double slotted foil in inserted into the bunch compressor. In the bunch compressor, there is a space-to-energy mapping, so the spatial windows spoil the bunch except in two energy regions, which are then the only regions able to lase. As energy maps to time in the undulators, this space-to-energy mapping becomes a space-to-time mapping for the emission. The result is reduced total brightness and emission confined to two short periods, i.e. pulses. The widths of the slits determine the widths of the pulses, and the slits' separation scales linearly with the delay \cite{Ding2015}. The space-to-energy mapping in the bunch compressor is equally important, and will jitter with the electron beam energy. Pulses up to 30 $\mu$J were produced in this way with photon energy centred close to 540 eV and separated by 14 eV. The repetition rate was 120 Hz, though complete pulse diagnostics operated at only 60 Hz. The temporal structure of the pulses is retrieved for the double slotted foil method using XTCAV.\\

\noindent{} In general, we find higher input-output correlation for the data from \cite{Sanchez-Gonzalez}. Testing on the highly correlated data with limited non-linearity helps to benchmark our theoretical prediction models.

\section{Predicting central photon energies with two-pulse data from experiment in [12]} 
\begin{figure}[ht!] 	
	\centering              
	\includegraphics[width=0.8\linewidth]{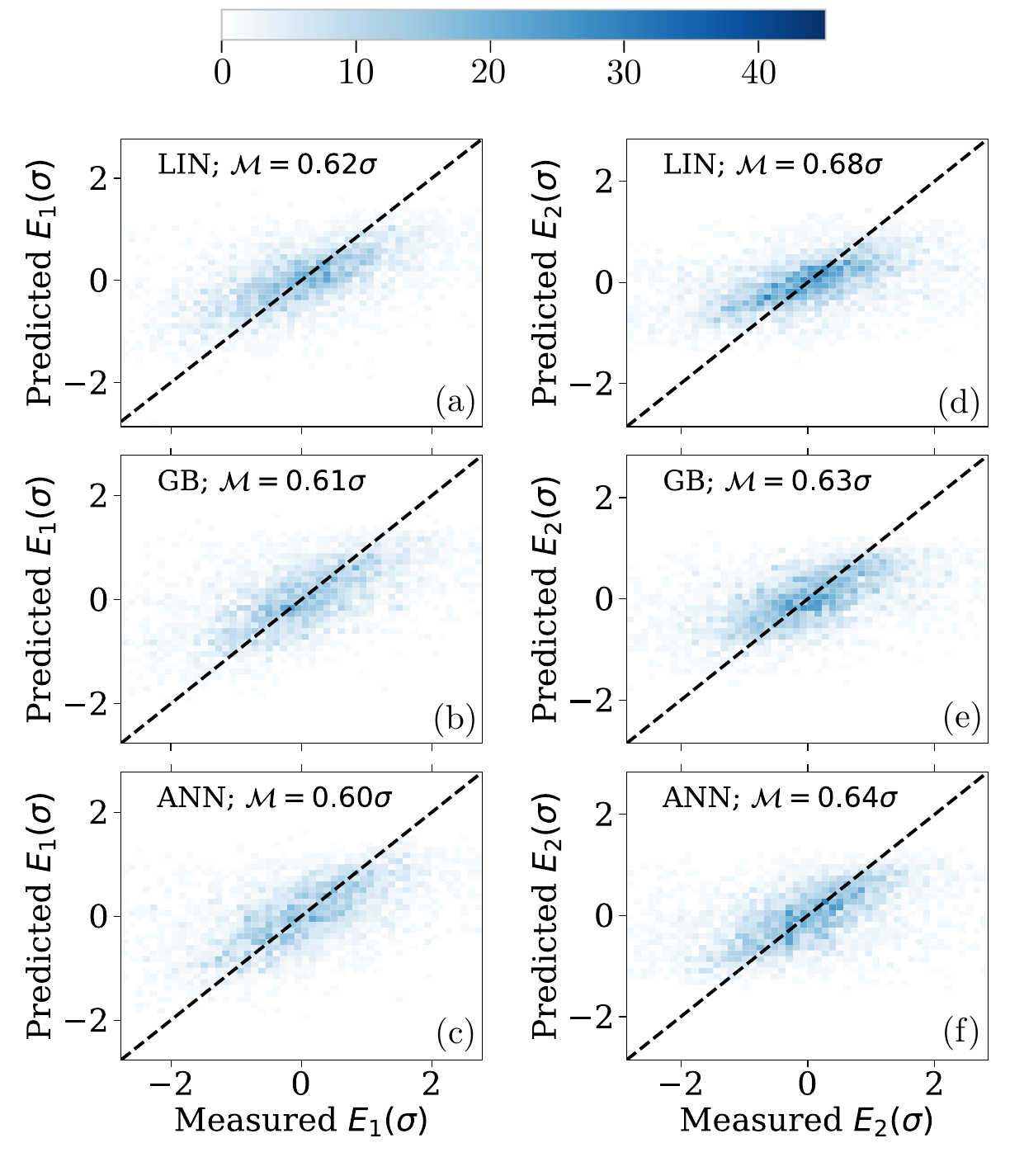} 
	\caption{Prediction of XFEL energies of the individual pulses for two-colour data: (a--c) compare the measured values of energies $E_1$ with the values predicted  by different ML methods, while (d--f) is the same for $E_2$. Top row panels represent linear regression model (LIN), middle row panels represent gradient boosting method (GB) and bottom row panels represent neural networks (ANN). The 2D histogram plots are constructed by grouping the data into 50 bins along each direction, where the density is indicated by the intensity of the blue colouring.} 
	\label{FigSupp1_eenergy}
\end{figure}

Figure~\ref{FigSupp1_eenergy} shows the validity of predicting central photon energies of the individual pulses ($E_1$ and $E_2$) using different machine learning methods. Despite the complex inter-dependence of these energies on the diagnostics, which is in some cases highly non-linear, the linear regression model (LIN) makes reasonable predictions of the central photon energy for either pulse, as seen in Fig.~\ref{FigSupp1_eenergy}(a) and (d).

\noindent{}Both gradient boosting (GB) and artificial neural networks (ANN) make better predictions than the  LIN models for the central photon energies of the individual pulses as depicted in Fig.~\ref{FigSupp1_eenergy}(b), (e) and Fig.~\ref{FigSupp1_eenergy}(c), (f) respectively. In general, independent of the prediction model, the mean absolute error for the predictions of $E_2$ are 2.4 times larger than for $E_1$ which can be attributed to the fact that Pulse 2 is the second harmonic of Pulse 1. Thus, the second pulse will experience effects of electron bunch energy jitter twice as much compared to Pulse 1 while the remaining difference may be attributed to the error in our energy measurements. It is promising to find that the GB model and the ANN model have similar accuracy in their predictions, especially since we find the GB models are faster to train compared to ANN models, at least by a factor of three.

\newpage
\section{Predictions of central photon energy and time delay for two-colour data from experiment in [23]} 

\begin{figure}[ht!] 	
\centering   
\includegraphics[width=0.8\linewidth]{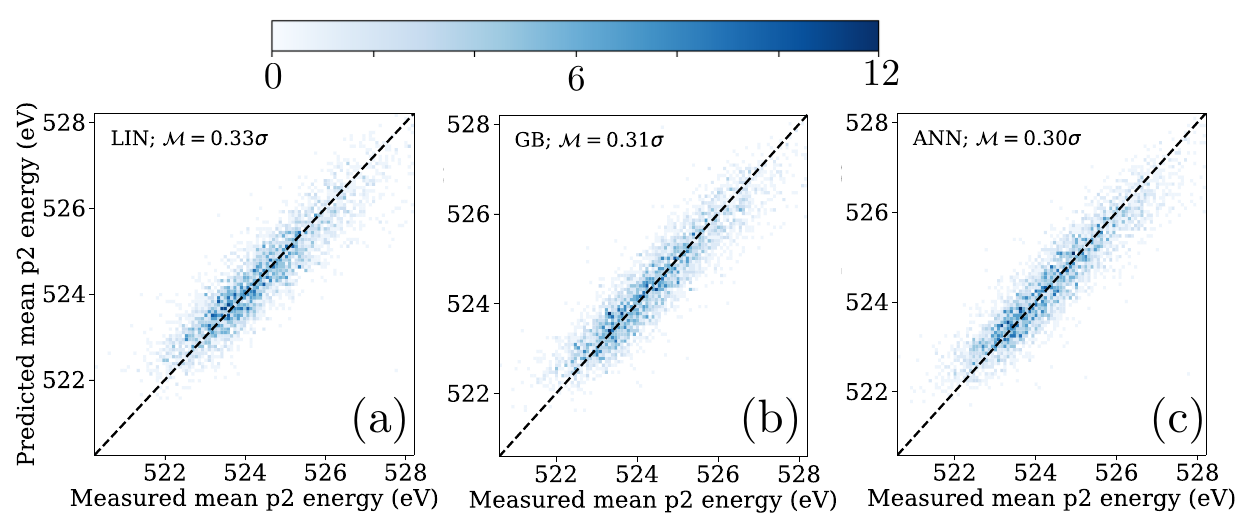}            
\caption{Comparing the measured values of the higher energy pulse from the two-colour data of an older operation mode \cite{Sanchez-Gonzalez} using the prediction from different ML methods: (a) linear regression model (LIN), (b) gradient boosting method (GB) and (c) neural networks (ANN). 2D histogram plots are constructed in similar way as Fig.~\ref{FigSupp1_eenergy}. Note that the energies shown here differ from those presented in  \cite{Sanchez-Gonzalez}  by a small offset owing to a scaling factor but doesn't affect the performance of the fit.} 
\label{FigSupp2_energy}   
\end{figure}

\begin{figure}[ht!] 	
	\centering             
	\includegraphics[width=0.8\linewidth, clip, trim={420, 250, 400, 250}]{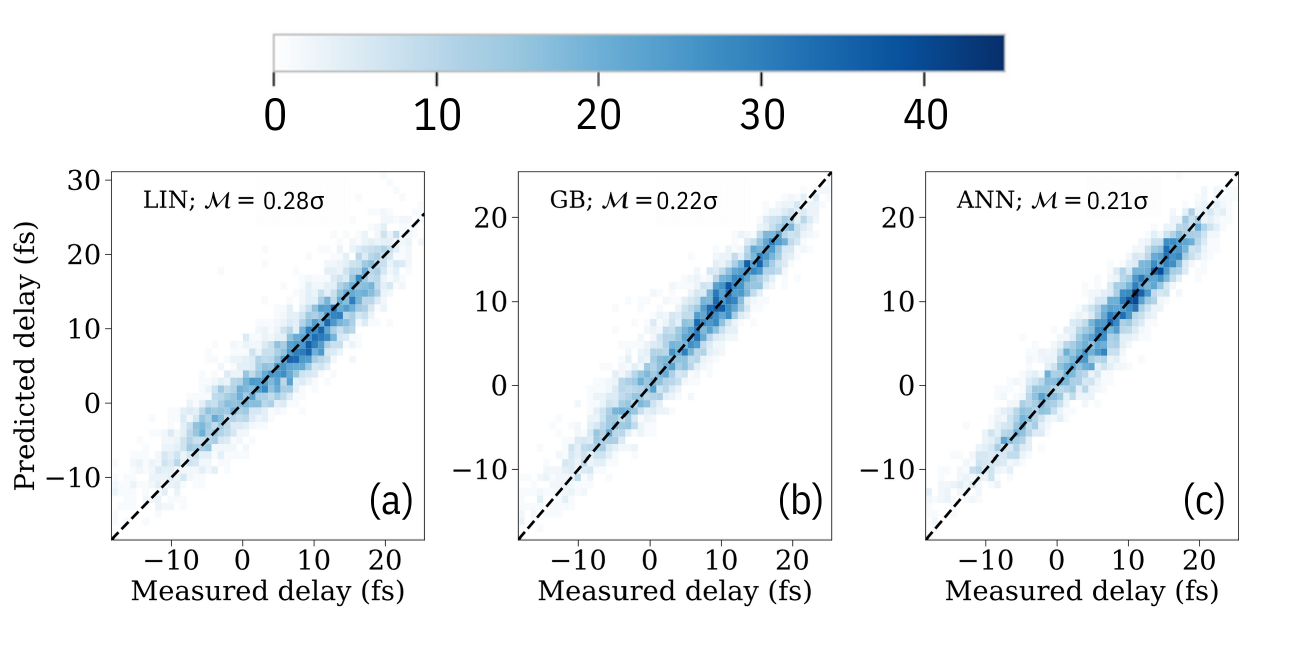} 
	\caption{Comparing the measured values of time delay for the two-colour data \cite{Sanchez-Gonzalez} using the prediction from different ML methods: (a) linear regression model (LIN), (b) gradient boosting method (GB) and (c) neural networks (ANN). The $\mathcal{M}$ values correspond to 2.39 fs, 1.87 fs, and 1.77 fs for LIN, GB, and ANN, respectively.} 
	\label{FigSupp3_delay}   
\end{figure}

Figures~\ref{FigSupp2_energy} and \ref{FigSupp3_delay} depicts the prediction of the central photon energy of the second pulse and time delay between the two pulses using the machine learning methods as used in the main manuscript. While the results agree with \cite{Sanchez-Gonzalez}, it is more efficient in training time due to reduced input parameter space, which is the result of our feature analysis. Interestingly, for this data,  the linear model here is sufficient to make accurate predictions with mean absolute error that is a fraction of the variance of the data. This perhaps can be understood by considering how the double-slotted foil affects the electron bunch used to create the pulses. In summary, the data in \cite{Sanchez-Gonzalez} seems to be less nonlinear in nature with high input-output correlation compared to \cite{Duris}.

\bibliography{Article_Latest/ML_XFEL_ARXIV}

\end{document}